\documentclass[12pt]{article}
\usepackage{graphicx}
\small

\begin{document}

\centerline{\large A possible solution for the faint young Sun paradox: Clues from the exoplanetary data}

\vskip 0.3cm
\centerline{\em Shashanka R. Gurumath$^{1}$, K. M. Hiremath$^{2,3}$, V. Ramasubramanian$^{4}$
and  Kinsuk Acharyya$^{5}$}

1. Department of Physics, Bengaluru City University, Bengaluru-560001

2. Indian Institute of Astrophysics, Bengaluru-560034, India; E-mail:hiremath@iiap.res.in

3. \#23, Mathru Pithru Krupa, 2nd Cross, 1st Main, BDA Layout, Bikasipura, BSK V Stage, Bengaluru-560111, India

4. Vellore Institute of Technology (VIT), Vellore, India

5. Physical Research Laboratory (PRL), Ahmedabad, India
\vskip 0.3cm
\begin{abstract}
Faint young Sun paradox (FYSP) is one of the interesting problems in solar physics. The present 
study aims to get a possible solution for the FYSP through sun-like G stars and their exoplanetary 
systems. Using physical properties of exoplanetary data, an empirical relationship between the 
rate of mass loss ($\frac{dM}{dt}$) with stellar mass (M$_{\star}$) and age ({\em t}) is obtained. We 
found mass loss rate varies with stellar mass as $\propto$ $(M_{\star}/M_\odot)^{-3.788}$ and proportional 
to the age as $\propto$ t$^{-1.25}$, 
which indicates rate of mass loss is higher during early evolutionary stages. Then we applied mass loss corrections 
to stellar masses of G-type stars with planets and obtained their initial masses at the early evolutionary stages.
Subsequently, we applied these relationships to calculate the mass loss rate and mass of Sun at the early evolutionar 
stage, which is found to be $\sim$ 10$^{-11}$ solar mass per year and $\sim$ (1.061$\pm$0.006) solar mass respectively. 
The higher solar mass can probably alleviate the problem of the faint young Sun paradox. Then the estimated initial 
stellar masses of the host stars are used to obtain a best power law relationship with the planetary masses that 
supports the hypothesis that the {\em massive stars harbour massive planets.} Finally, by using the same empirical 
power law, planetary mass in the vicinity of Sun is estimated to be $\sim$ (0.84$\pm$0.19) Jupiter mass, which is 
much higher compared to the present solar terrestrial planetary mass. Hence, this study also suggests that there is 
a missing planetary mass in the vicinity of the Sun, which can solve the FYSP problem.

\end{abstract}

\section{Introduction} 
The solar system originated from the gravitational collapse of dense interstellar cloud known as the solar 
nebula which primarily consisted of gas and dust particles (Woolfson, 2014). 
Many theories are proposed to understand the dynamical 
evolution of Solar system planets (Nesvorny, 2018) viz., the formation of 
Jovian planets, the low mass of terrestrial planets by 
considering migration scenario (Tsiganis et.al. 2005, Walsh et.al. 2011) 
and, the higher density of the solar nebula
(Alibert et.al. 2005, Crida 2009, Mordasini et.al. 2012) during the early history of solar 
system formation. However, many questions remain unresolved. Faint young Sun paradox (FYSP) is one of 
such question (Sagan \& Mullen 1972, Feulner 2012) which requires 
a suitable explanation for liquid water on the Earth during the early evolutionary stages. 

Several ideas are proposed to solve the FYSP, viz., 
deviation from the standard solar model (Gaidos et.al. 2000, Shaviv 2003)
increase the fraction of greenhouse gases in the early Earth/Mars 
(Goldblatt \& Zahnle 2011, Feulner 2012) etc. One of the promising 
solutions for FYSP is Sun's mass during an early stage 
of evolution must be slightly higher (1.03 - 1.07 $M_{\odot}$, 
where $M_{\odot}$ is the mass of the Sun; 
Section 7.1.2 of Gudel 2007, Minton \& Malhotra 2007) 
compared to the present mass, which can provide a higher level of 
solar luminosity resulting in more energy output than was predicted. 

Present day mass loss rate of Sun is found to be $\sim$ 10$^{-14} M_{\odot}$ /yr, 
which is estimated only from solar wind. If one estimates initial Sun's mass by 
assuming this mass loss rate is constant, it results in increase of 0.05$\%$
solar mass that yields negligible increase in Sun's luminosity. 
Wood et.al. (2002) obtained a non-linear relationship between mass loss rate 
and stellar age that suggests mass loss is high during early stages of 
star formation. In addition, theoretical studies 
(Drake et.al. 2013 and references there in, Cranmer \& Saar 2011)
also suggest that Sun mass loss rate is higher ($\sim$ 10$^{-11} M_{\odot}/yr$) during its early evolutionary stages in
order to have liquid water on Earth.

Discovery of exoplanets promted scientific community to understand planet formation in a larger broader scale and 
compare physical and orbital characteristics (Winn \& Fabrycky, 2015) with the Solar system planets.
A reasonable subset of exoplanets are around the sun-like G-type stars, therefore in the present study an attemps has beed made to 
understand Faint young Sun paradox by studying the exoplanetary data. We also examined whether present terrestrial mass in the vicinity 
of the Sun is compatible with mass of exoplanets in the vicinity of their host stars. With these aims in mind, section 2 presents 
the data and analysis. Section 3 describes the methods to estimate the rate of mass loss of the host stars from the observed data set and, from the empirical 
rate of mass loss relationship (Jager \& Nieuwenhuijzen, 1988). 
The results and conclusions are presented in section 4.

\section{Data and Analysis}
\label{sect:DA}

For the present study, physical and orbital characteristics of 
Sun-like G stars and their exoplanets are considered 
from the website {\em http://exoplanet.eu/catalog/}. 
The Sun-like G-type stars in the present study are defined 
based on the spectral G type provided in the above-mentioned website.
Further, we impose the following constraints on the physical 
properties of stars and exoplanets:
(i) consider the host star that has $\le 50\% $ error in its age,
(ii) in order to avoid the ambiguity in considering the brown dwarf candidates, 
restrict the maximum planetary mass limit to 13 $M_J$ 
(where $M_{J}$ is mass of the Jupiter) and,
(iii) avoid the binary stellar systems.
With these constraints, we left with 114 host stars that have 147 exoplanets.
Among 114 host stars, 17 are multi-planetary hosts with 49 planets.
The relevant data is given in the Table \ref{T1}. 
First column represents the name of an exoplanet. Second and third
columns indicate the planetary mass and its error in terms of 
Jupiter's mass respectively. Fourth and fifth columns subsequently describe 
the orbital distance of a planet and its error in terms of AU.
Sixth and seventh columns represent the stellar mass and its error 
in terms of Sun's mass respectively. Last three columns subsequently
indicate the spectral type, age (in Gyrs) and error in age.

One can argue that data bias ((i) high mass planets in the
vicinity of stars and (ii) high uncertainties in the stars ages)
can lead to misleading results. However, when one examines the 
data presented in Table \ref{T1}, following two 
important facts completely rule out such biases. 
For example, in addition to hot-Jupiters,
the data set also consists of many low-mass planets in the vicinity
of their host stars viz., Kepler-10 b, Kepler-11 b, HD 20794 b, 
HIP 68468 b, etc. In addition, majority of stars in the data set
have less than 30\% uncertainties in their estimated ages, 
which help in removing the inconsistencies that arise in estimating 
the stellar rate of mass loss. Further, one can also notice that the 
error bars in the stellar and planetary masses are small 
(on average 10\%). 
Hence, present data set is most appropriate to extract the useful
scientific results.

\section{Estimation of mass loss from the host stars}
\label{3}
The solar type stars lose their mass heavily 
when they were young (Linsky et.al., 2004) and,
the rate of mass loss decreases as the stars evolve 
(Wood et.al., 2005).
This is because, at later evolutionary stages of Sun-like stars, 
the rotation rate decreases (Hartmann 1985)
that results in less magnetic activity.
However, mass loss is small for massive stars ($\sim 7M_{\odot}$) 
in their main sequence stage and increases with the 
late stage of stellar evolution (Huang et.al. 1990).
Hence, stellar mass loss eventually affects the internal structure of a star. 
The rate of mass loss of host stars ranges from $10^{-14}$ to 
$10^{-4}$ $M_{\odot}$ /year (Redgway et.al. 2009)
depending upon the different stellar
properties viz., stellar mass, luminosity, radius, and temperature.
Since the stars are very active in their early evolutionary stages 
(Hiremath 2009 and references therein, Hiremath 2013),
there might have more chance of losing heavy mass from them, that might affect the
nearby planets.

\begin{figure}
	\centering
	{\includegraphics[width=15pc,height=15pc]{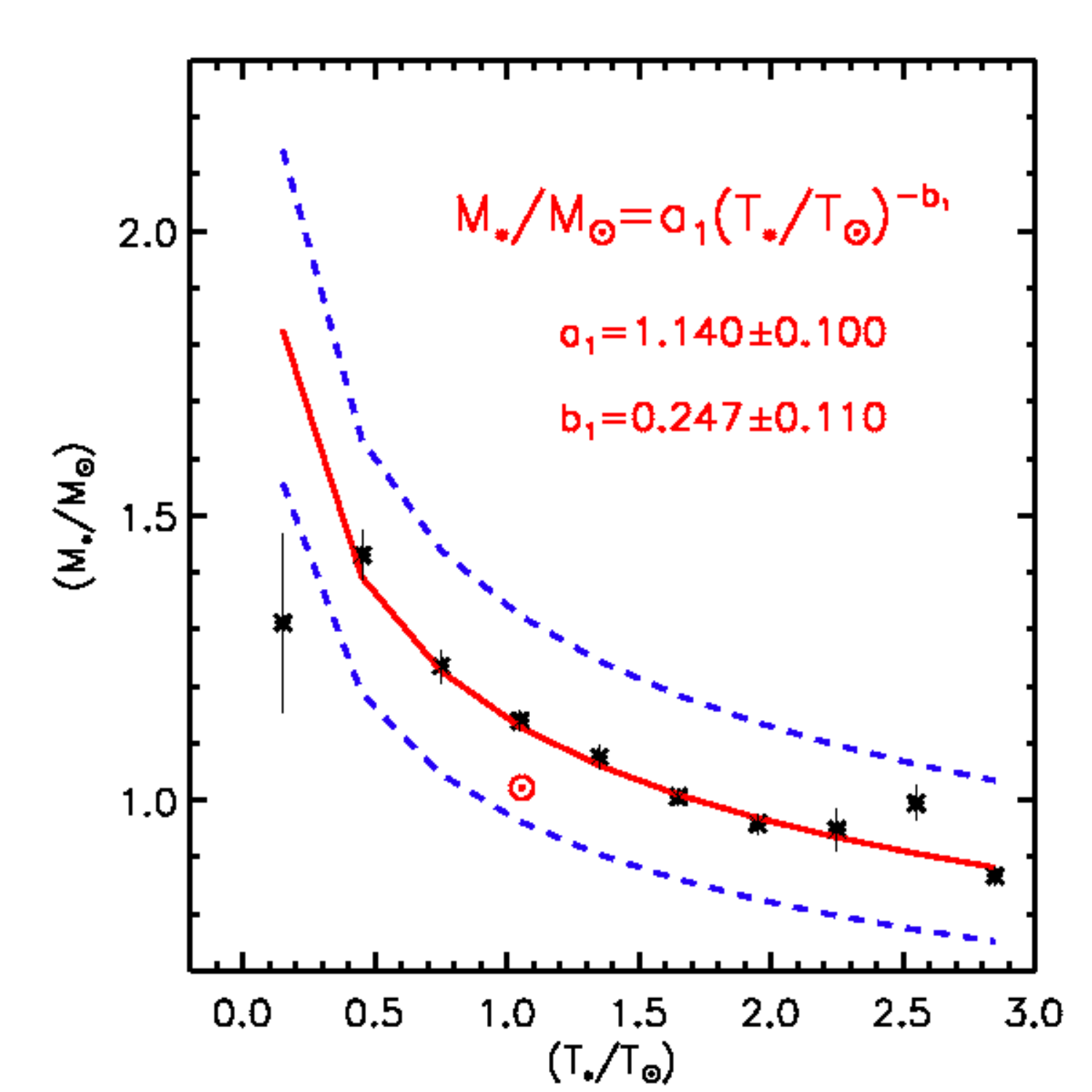}}
	\caption{Illustrates the dependence of stellar mass with its age. 
		The continuous line indicates the best power law fit between both the variables
		and the dashed lines indicate the one sigma error level.
		$\odot$ symbol indicates the position of the Sun.}
	\label{Fig1}
\end{figure}

\subsection{Rate of mass loss estimated from the host stars that have exoplanets}
\label{3.1}

Since we do not know whether the rate of mass loss is same for 
stars with and without exoplanets, first an attempt is made 
to understand the variation of stellar masses (that have exoplanets) with their ages.
Figure \ref{Fig1} (star's mass M$_{\star}$ and star's age T$_{\star}$ are 
normalized with the Sun's mass M$_{\odot}$ and age T$_{\odot}$ respectively) illustrates
a power law relationship between stellar mass and
age such that $\frac{M_{\star}}{M_{\odot}} \sim (\frac{T_{\star}}{T_{\odot}})^{-0.247}$.
In this figure, the x-axis is binned with a size of 
0.3 $\frac{T_{\star}}{T_{\odot}}$, and the y-axis represents
the average stellar mass in that bin with an error bar estimated 
from the ratio $\frac{\sigma}{\sqrt{n}}$ (where $\sigma$ is a
standard deviation and {\em n} is the number of data points in each bin). 
Such an empirical power law suggests that the stellar rate 
of mass loss is high during the initial evolutionary stages.

With the present data set that consists of stellar mass and age and, by
making use of numerical differentiation, the rate of mass loss 
$\frac{d(M_{\star}/M_{\odot})}{d(T_{\star}/T_{\odot})}$
for each star is estimated. 
Now onwards, $\frac{M_{\star}}{M_{\odot}}$ and $\frac{T_{\star}}{T_{\odot}}$ 
are represented as $M$ and {\em t} respectively.
Further, we find that estimated $\frac{dM}{dt}$
fits a power law relation with the stellar mass.
By making use of this power law relation, 
the estimated (see Table \ref{T2}, second column, second row)
rate of mass loss of the Sun is found to be $\sim 10^{-11}$ $M_{\odot}$/yr, 
which matches very well in the range of mass loss computed by
the previous studies (Drake et.al. 2013, Cranmer 2017) 
if one considers mass loss due to coronal mass ejections also.

\begin{figure}
	\centering
	\includegraphics[width=0.5\textwidth]{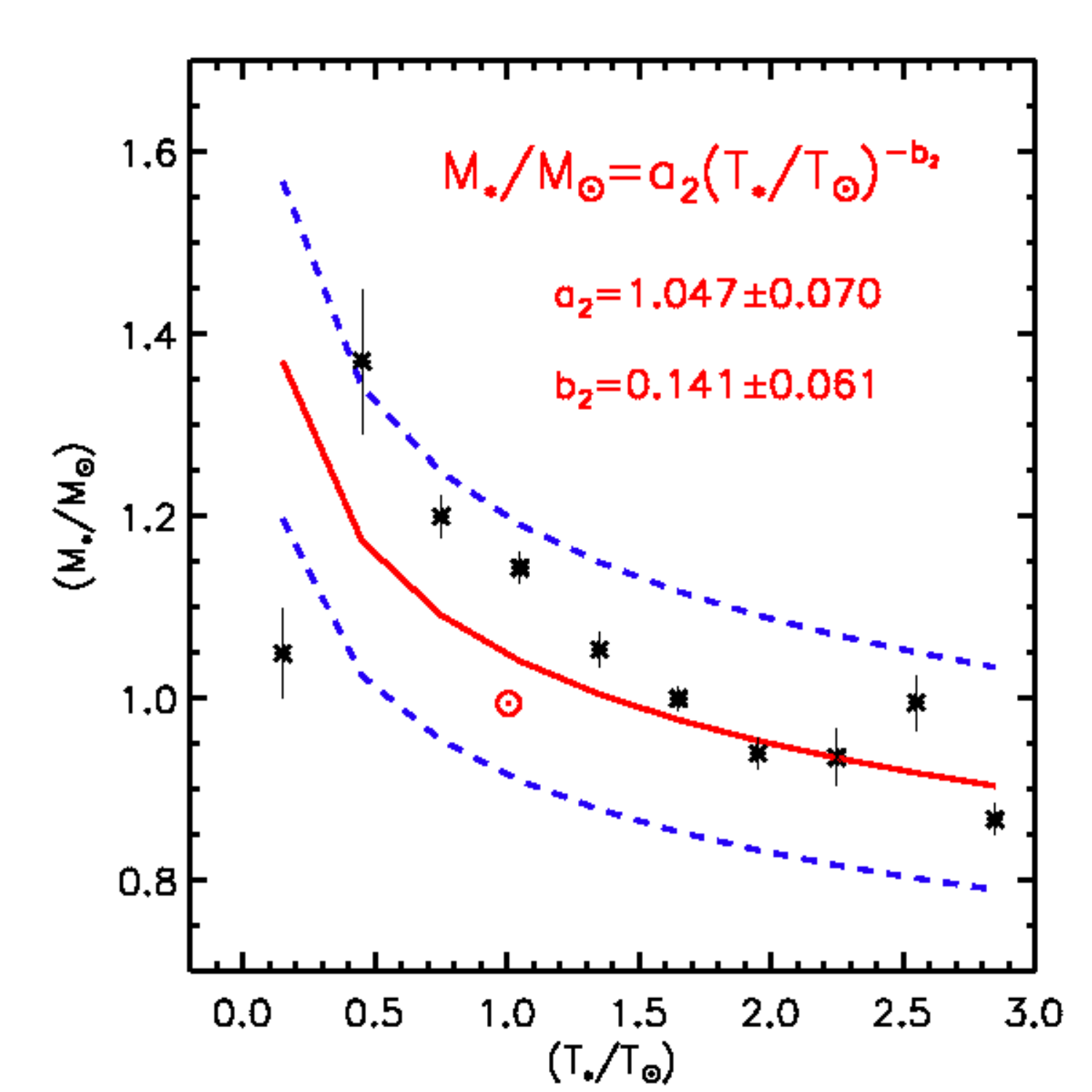}
	\caption{Illustrates the dependence of stellar mass with its age 
		for the dataset without giant stars. 
		The continuous line indicates the best power law fit between both the variables
		and the dashed lines indicate the one sigma error level.
		$\odot$ symbol indicates the position of the Sun.}
	\label{Fig2}
\end{figure}

Careful observation of Table \ref{T1} reveals that dataset
consists of few giant stars. Since these giant stars have the high rate of 
mass loss compared to other main sequence stars, one can argue that their presence
significantly contribute to the high rate of mass loss of the stars.
In order to confirm whether these giants have really affected the rate of 
mass loss, stellar mass versus age relationship without the giants is illustrated
in Figure \ref{Fig2}. Further, a relationship is examined between 
the rate of mass loss of stars (estimated as described above) and their
stellar masses that yields the rate of mas loss
of Sun to be $\sim$ 0.9 x 10$^{-12}$ M$_{\odot}$/yr
which is almost equal to the previously estimated value. 
Both these results strongly suggest that contribution due to
coronal mass ejections from the Sun substantially increases its 
present rate of mass loss that is estimated only from the solar wind.
One can also notice that Sun's position in both the Figures (\ref{Fig1} and \ref{Fig2})
is just below the fitted line within one sigma error level. 
That means, Sun's evolutionary path is no different than the 
stars evolutionary path that have exoplanets. Hence,
Sun might have experienced a major mass loss during its early evolutionary
stages that might have kept Earth and Mars atmosphere probably in warm conditions.

Since stellar ages have appreciable but not very high magnitude of error bars,
in order to validate the obtained law of mass loss, we have also verified 
from the data of 42 stars that have accurate stellar 
mass (average error in stellar mass is $\sim$ 3.07\%) 
and stellar age (average error in stellar age is $\sim$ 11.80\%) computed 
from the asteroseismic method (Metcalfe et.al. 2014). 
Figure 3 illustrates the stellar mass versus age relationship
for the stars (with no detected exoplanets) whose properties
are accurately determined from the asteroseismic method.
The x and y axes are binned as described previous paragraph.
Careful observation of Figures \ref{Fig1} - \ref{Fig3}
reveals that, during early stellar age ($\le$ 0.3 $\frac{T_{\star}}{T_{\odot}}$),
stars with exoplanets experienced a high rate of mass loss compared to
stars without exoplanets.
However, cause for the excessive rate of mass loss from the stars
with exoplanets during early evolutionary stage is beyond the scope of this study.

\begin{figure}
	\centering
	\includegraphics[width=0.5\textwidth]{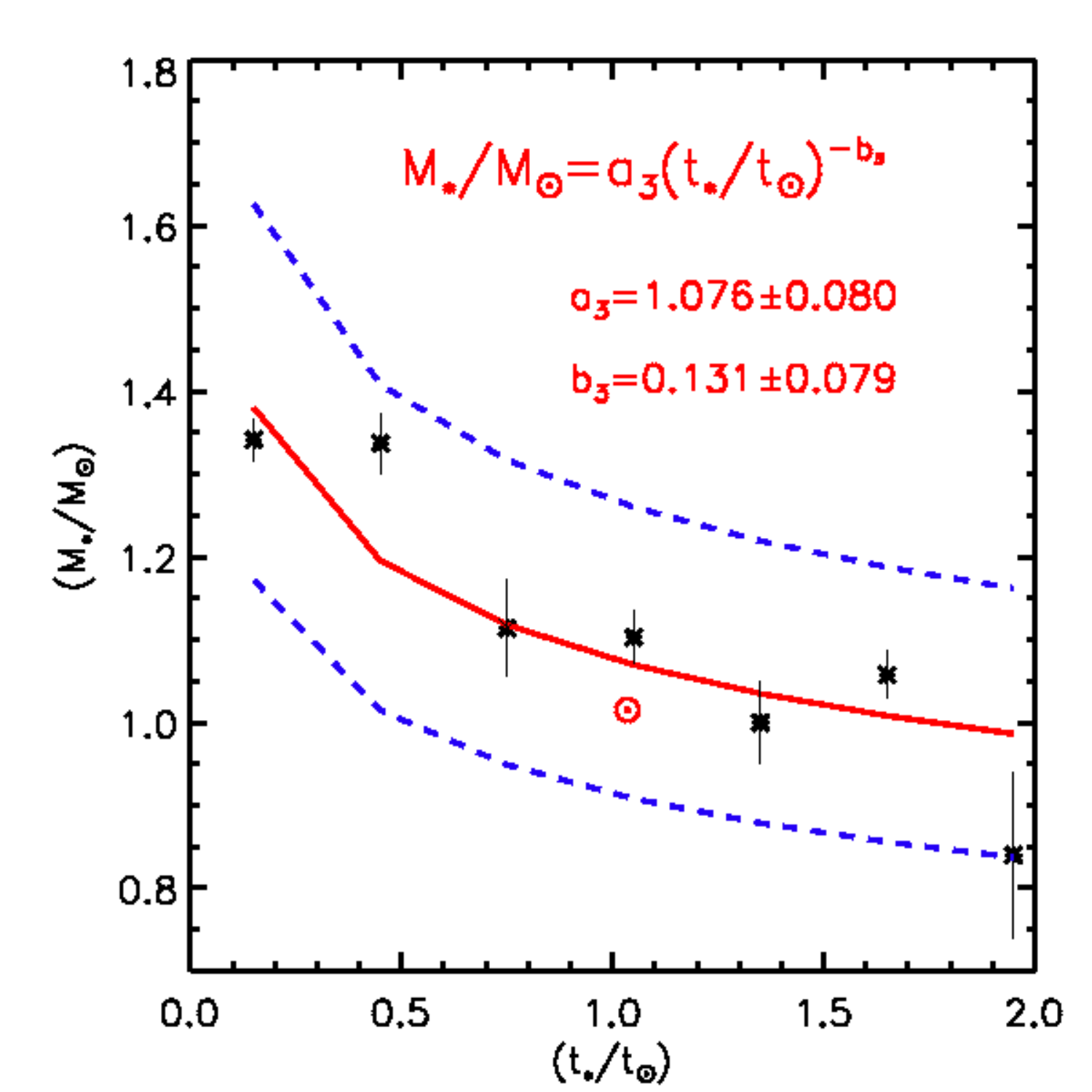}
	\caption{Illustrates the dependence of stellar mass with its age 
		for the stars whose properties are estimated from the asteroseismic method.
		The continuous line indicates the best power law fit between both the variables
		and the dashed lines indicate the one sigma error level.}
	\label{Fig3}
\end{figure}

\subsection{Rate of mass loss estimated from the observations of stars}
\label{3.2}
There is every possibility that, a mismatch between the estimated and observed
rate of mass loss of the Sun could be due to not using the observed rate of
mass loss of host stars.
Hence, the observed (direct estimation) rate of mass loss of host stars,
irrespective of whether stars harbour planets or not, 
are considered from (Cranmer \& Saar, 2011).
With this data, a power law relationship is obtained 
between the rate of mass loss of stars and their masses.
Using this power law relationship, the present rate of mass loss of 
the Sun is estimated to be $\sim$ 0.8 x $10^{-12} M_{\odot}$/yr 
which is of the same order ($\sim$ $10^{-11} M_{\odot}$/yr) as 
estimated from the host star mass-age relationship.
For additional check, an empirical rate of mass loss relationship
given by \cite{Jager} is also used to estimate the Sun$^{\prime}$s present
rate of mass loss which is found to be of 
similar order ($\sim$ $10^{-14} M_{\odot}$/yr; see also Wood et.al. 2005).
Obvious reason for the difference of rate of mass loss estimated
by Wood et.al. (2005) and our estimation is that, (Wood et.al. 2005) 
estimate the rate of mass loss due to stellar wind only.

All these three empirical rate of mass loss laws and
estimated present Sun$^{\prime}$s rate of mass loss are presented in
the first two rows of Table \ref{T2}. It is interesting to note that
all the three empirical laws suggest
power law relationships between the rate of mass loss of stars and their respective masses. 
Hence, it is obvious from these relationships that,
high mass stars have a high rate of mass loss.

One can also notice from Table \ref{T2} that
different mass loss laws yield different initial Sun's mass. 
Among all the mass loss laws, the estimated uncertainties 
in the exponents of power laws are least for the mass loss law
that is estimated from the host stars data (second column, Table \ref{T2}).
In addition, the Sun's rate of mass loss estimated from this law very 
well matches with the rate of mass loss estimated by (Drake et.al. 2013). 
Hence, among these three relationships, best relationship is the one 
which is derived from the host stars data as described in section 3.1.

\subsection{Rate of mass loss as a function of age}

Previous study suggests that rotation rate and magnetic activities of stars,
that are main reason for mass loss, decrease with stellar age (Vidotto et.al. 2014).
Due to high magnetic activity, heavy mass loss occurs in early evolutionary stages of a star.
Wood et.al. (2002) theoretically obtained a relation that suggests stellar mass loss
decreases with its age. In the present study, by making use of estimated
rate of mass loss of stars with exoplanets (as described in previous sections) 
and their ages, we also obtained a non-linear relationship such that

\begin{equation}
\frac{dM}{dt} = \alpha T_{\star}^{\beta},
\end{equation}

\noindent where $\alpha$ = (0.286$\pm$0.007), $\beta$ = (1.253$\pm$0.012)
and stellar age $T_{\star}$ is in units of Gyrs.
Figure 4 illustrates the rate of mass loss as a function of stellar age
as given by equation (1). This figure also indicates that mass loss rate is high
during early evolutionary phases of star and it decreases with age.
We compared our result with the result by Wood et.al. (2002) with different 
intercepts as mentioned in Figure caption. 
It shows that power law  (equation 1) from the
present study fits the data better than the previous study.
Furthermore, by using equation (1), 
rate of mass loss of the Sun is found to be
$\sim$ 2.54 $\times$ 10$^{-11} M_{\odot}/yr$ 
during its early evolutionary stages. 
Hence, this analysis also supports the 
results that are obtained in section 3.1. 

\begin{figure}
	\centering
	\includegraphics[width=0.5\textwidth]{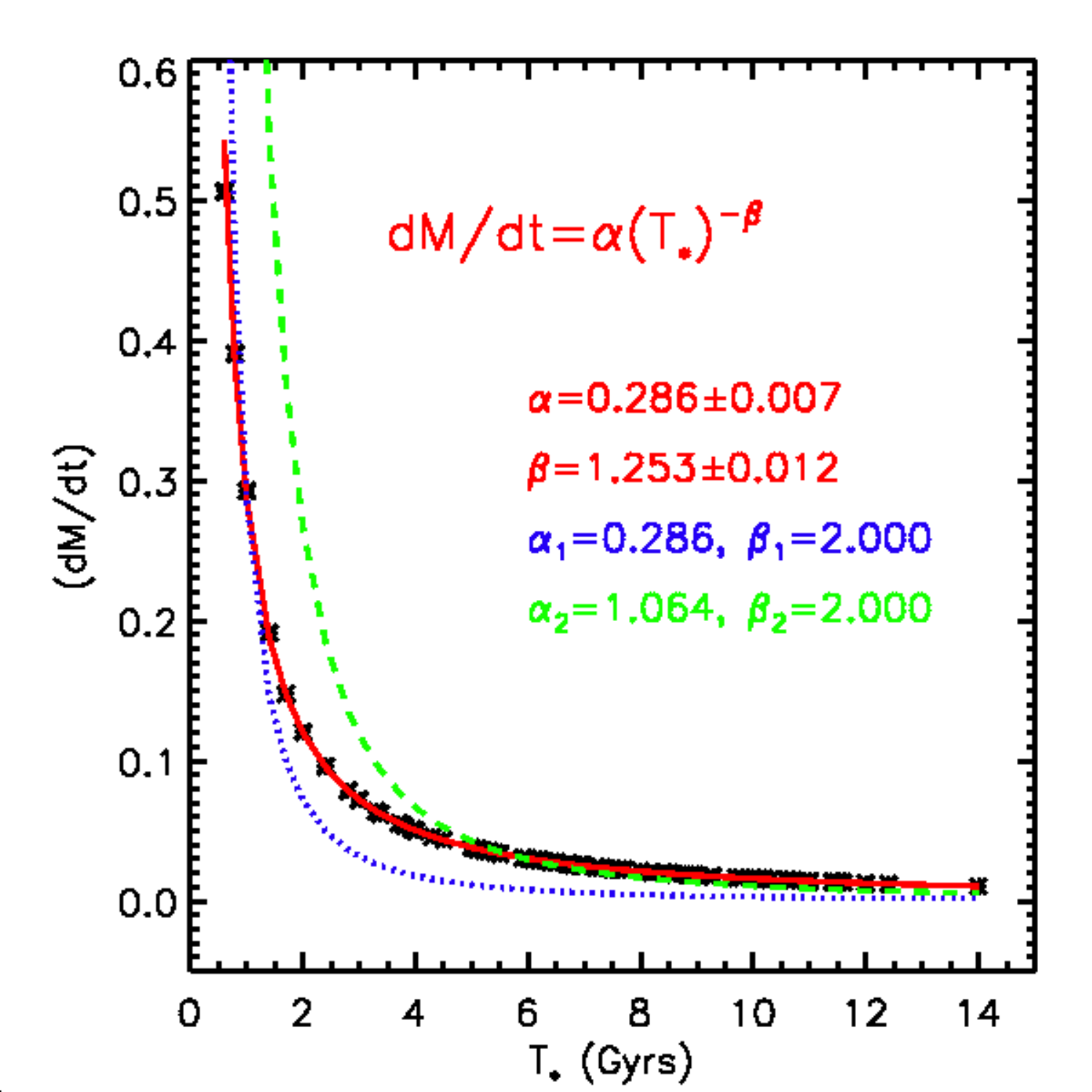}
	\caption{Illustrates the variation of mass loss rate with respect to stellar age.
		The solid line indicates the best power law fit from the present study, 
		dashed line indicates fit by using exponent from Wood et.al. (2002) and,
		dotted line indicates fit by exponent from Wood et.al. (2002) and intercept from present study.}
	\label{Fig3}
\end{figure}

\subsection{Computation of initial stellar mass}
With these three empirical power laws (as presented
in the first row of Table \ref{T2}) of rate of mass loss, 
initial stellar mass $M_{ini}$ can be computed in the following way

\begin{equation}
M_{ini} = M_{pre} + \int_{t_{1}}^{t_{2}} \mathrm{d M \over d t} \mathrm{d}t,
\end{equation}

\noindent where $M_{pre}$ is present stellar mass,
$t_{1}$ and $t_{2}$ are initial and present ages of a star respectively. 

On the right-hand side (RHS) of equation (2), the second term represents the total 
mass lost by a star from its initial age $t_{1}$ to present age $t_{2}$. 
Hence, by adding total lost mass to present mass, initial 
mass $M_{ini}$ of a star can be estimated. In the present study, 
total mass lost (numerical integration of the second 
term in RHS is performed) from the host stars
is estimated and is added to the present mass $M_{pre}$ in 
order to get the initial mass $M_{ini}$.
For estimating total rate of mass loss, initial ages of stars 
are assigned to be $\sim$ 50, 60, 70, 80, 90, 100 and 200 million years
and, initial stellar mass is estimated for these ages.
However, among all these assigned initial stellar ages, the estimated initial stellar
masses converge and saturate at 50 million years than the other assigned ages.
Hence, we accept 50 million years as the initial age of a star.

\section{Results}

\begin{figure}
	\centering
	\includegraphics[width=0.5\textwidth]{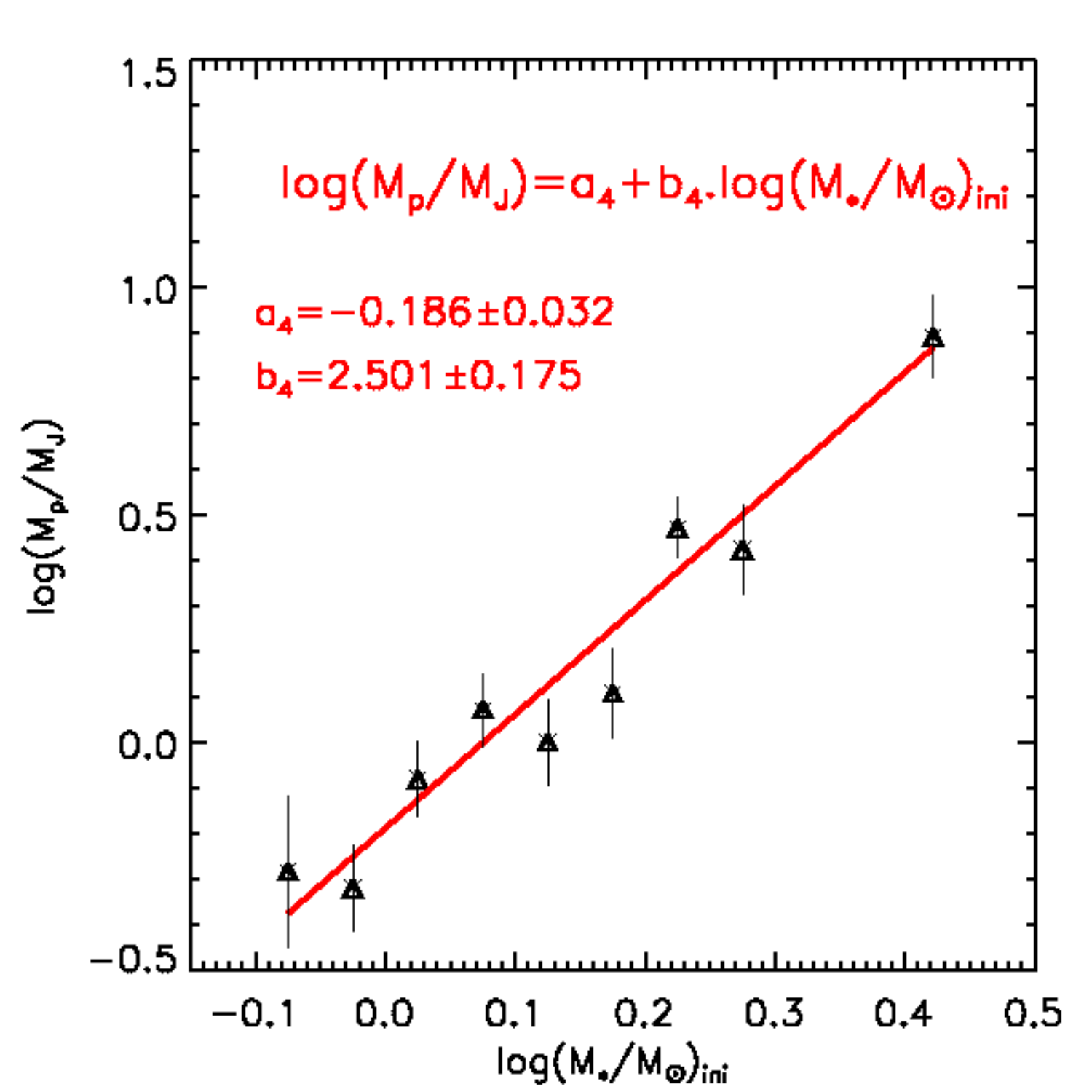}
	\caption{Illustrates the dependence of present mass of the exoplanets
		with the initial mass of star.
		The continuous line indicate the best fit line between
		the  planetary mass and initial stellar mass and dashed lines indicate
		the one sigma error level.}
	\label{Fig4}
\end{figure}

After computing the initial stellar masses of the host stars, 
association between the initial stellar
masses and the planetary masses is examined. In the case of 
multi-planetary systems, all the planetary
mass is added that gives the information about the total 
amount of planetary mass for respective host star.
As illustrated in Figure \ref{Fig4} we find that 
these two parameters are non-linearly (power law)
dependent on each other. In Table \ref{T3}, three power laws 
that relate the initial host star mass with the
planetary mass are presented. In each case, to estimate 
the initial stellar mass, separate rate of mass
loss correction is applied as mentioned in the first column of Table \ref{T3}.

The power law between the initial stellar mass and planetary 
mass as illustrated in the Figure~\ref{Fig4}
suggests that, {\em the massive stars harbor massive planets in their vicinity}.
These results are also consistent with the previous studies of observational 
(Laws et.al. 2003, Johnson et.al. 2007, Lovis \& Mayor 2007, Johnson et.al. 2010)
and theoretical inferences (Ida \& Lin, 2005).
For different spectral type, Lovis \& Mayor (2007) present 
a relationship between exoplanetary mass and stellar mass.
However, the present study is different than the previous studies as we have consistently 
applied the mass loss correction to the stellar mass and also quantitatively 
estimated the relationship between the stellar mass and the planetary mass.

\subsection{Estimation of initial mass of the Sun}
Although it is not proper to compare physics of
the Solar system with physics of exoplanetary systems,
since the Sun has similar physical characteristics 
such as mass, it is interesting to estimate Sun's lost mass and 
initial mass from the rate of mass loss relationships 
(see Table \ref{T2}, second row). The estimated rate of 
mass loss of the Sun from the de Jager's empirical formula is
no different than the present Sun$^{\prime}$s rate of mass loss, that is 
estimated only from the solar wind. This value almost 
matches with the rate of mass loss estimated by Wood et.al. (2005) if 
one considers the contribution due to solar wind only.
However, from the other two relations, indeed the Sun has an excess rate of mass loss.
Especially, if one accepts the rate of mass loss due to coronal mass 
ejections (Drake et.al. 2013, Cranmer 2017) also, then 
the rate of mass loss of the Sun increases by $\sim 10^{3}$ times
compared to a rate of mass loss estimated from the solar wind alone 
and hence inevitably Sun must have an excess initial mass. 
As we have already mentioned in 
section 3.2, the best relation between the rate of mass loss 
and stellar mass is one which is obtained from the host stars data. 
Hence, by using rate of mass loss (Table \ref{T2}, first row and second column)
from host stars data we 
have estimated the Sun$^{\prime}$s initial mass 
which is found to be $\sim (1.061\pm0.006) M_{\odot}$.
In addition, by making use of equation (1),
the Sun's initial mass is estimated to be
$\sim (1.05\pm0.01) M_{\odot}$.

It is to be noted that our estimated initial mass is consistent with the 
initial mass as estimated in the previous studies
(Whitmire et.al. 1995, Section 7.1.2 of Gudel 2007).
Moreover, the cumulative mass lost by the Sun during its evolution
until present age is 0.061 M$_{\odot}$ which is about 1\% of the 
solar mass as recently predicted by Cranmer (2017).
Hence, during early evolutionary stages, indeed
Sun's mass is slightly higher than the present mass. 
It is also to be noted that this estimated excess initial mass 
of the Sun is in the range (1.03 - 1.07 $M_{\odot}$; 
section 7.1.2 of Gudel 2007) of Sun's initial mass that requires Earth
to retain the liquid water.
This astrophysical solution is a possible reason for 
alleviating the long-standing problem of faint young Sun paradox.

\subsection{Estimation of initial planetary mass in the vicinity of Sun} 
With the calculated Sun$^{\prime}$s initial mass and making use of  
empirical relationships as presented in Table \ref{T3}, 
estimated average planetary mass in the vicinity of 
Sun is found to be $\sim (0.84\pm0.18) M_{J}$.
In contrast, the present total planetary mass (that includes masses
of all the terrestrial planets and asteroids) in the vicinity of Sun is 
$\sim$ 0.006 $M_{J}$ ($\sim$ 2 earth mass).
One can notice that the difference between estimated and 
observed planetary mass of the terrestrial planets suggests that
a substantial amount of planetary mass is missing in the vicinity of Sun. 
Careful observation of stars within the spectral type G provides 
an evidence of huge ($\sim$ 0.8 M$_J$) planetary mass in their vicinity.
Further, if we concentrate on stars that belong to spectral type G2 (similar to Sun),
the average planetary mass within their vicinity is found to be $\sim$ 1.3 $M_J$.
In addition, there are eight stars in the present dataset 
with similar mass ($\pm$ 0.2 $M_{\odot}$, along with spectral type G2) as that of the Sun
whose average planetary mass within their vicinity is $\sim$ 0.7 $M_J$.
Among these eight stars, minimum planetary mass within vicinity is $\sim$
0.3 $M_J$ for the star WASP-126 and, maximum planetary mass is $\sim$ 1.1 $M_J$
for the star WASP-95. These all evidences suggest that the Sun is in deficiency 
of planetary mass within its vicinity.

Probably during early phase of the Solar system evolution, 
some part of that missing mass might have 
accreted onto the Sun like other stars (Matt et.al. 2012) or some part might have 
blown off to space due to intense solar radiation 
(X-rays, EUV radiations) or protoplanetary disk might 
have blown off by photoevaporation (Mitchell \& Stewart, 2010)
or part of this missing mass might have 
migrated outwards (Walsh et.al. 2011) from the
vicinity of Sun during the early history of Solar system formation. 
Another possibility could be due to dynamical barriers imposed by 
the gas giants (Izidoro et.al. 2015) for migration of 
super earths, although Kepler data do not support 
such a migration scenario (Chatterjee \& Tan 2015).
Hence, further clear understanding of deficiency 
in the present planetary mass in the vicinity of Sun 
from the perspectives of theory and analysis of the 
observed physical characteristics of exoplanetary
data is necessary.

Since the Sun-like stars were very active during early evolutionary stages 
(Wood et.al. 2005, Hiremath 2009, Hiremath 2013, 
combined mass loss due to solar wind
and coronal mass ejections might have affected the mass of the nearby planets. 
As for estimated initial planetary mass $(0.84\pm0.18) M_{J}$ in the vicinity of Sun, 
since the planets also lose mass due to intense radiation from 
the host stars, with a caveat we have to conclude that, 
unambiguous planetary mass loss correction is also necessary. 
Presently, the planetary mass loss is not well understood and 
mass loss correction for the planets is
beyond the scope of this study. However, in case mass loss correction for the 
planets is also applied, estimated initial planetary mass 
might further increases and aggravates the deficiency of
planetary mass in the vicinity of Sun.

\section{Conclusions}
To conclude this study, physical and orbital characteristics of 
Sun-like G stars and their exoplanets are used to seek a
possible solution for the faint young Sun paradox.
Having a dataset of sun-like stars distributed over a 
range of stellar ages, a relationship between stellar mass and age
is obtained that suggests mass stellar mass decreases with
stellar age. By using this relationship and having knowledge of
stellar age, rate of mass loss of each star is estimated.
By estimating the rate of mass loss of each star,

\begin{itemize}
\item we obtained a non-linear relationship between the 
rate of mass loss of host star and its mass such that
$(dM/dt) \propto (M_{\star}/M_{\odot})^{3.79}$.

\item We obtained a relationship between 
mass loss of a star and its age such that 
$(dM/dt) \propto (T_{\star}/T_{\odot})^{-1.25}$ 
that suggests, stars experience high mass loss rate during
their early evolutionary stages.

\item Rate of mass loss of the Sun during early evolutionary stage is estimated to be $\sim$ 10$^{-11} M_{\odot}$/yr.
This high value of rate of mass loss can keep liquid water on Earth,
that is also suggested by previous studies 
(Cranmer 2017, Minton \& Malhotra 2007, Whitmire et.al. 1995).

\item By using Sun's rate of mass loss its initial mass 
is estimated to be $\sim(1.061\pm0.006) M_{\odot}$,
that can alleviate the faint young Sun paradox.

\item We acquire a non-linear relationship between initial stellar mass
and planetary mass that suggests massive stars harbour massive planets.

\item From the relationship between initial stellar mass and their exoplanetary mass, 
initial planetary mass in the vicinity of Sun is estimated to be $\sim (0.84\pm0.18) M_{J}$, 
in excess of planetary mass concentrated at the present epoch. 
The excess solar planetary mass lost is conjectured with 
different views. 
\end{itemize}

\begin{table*}
	\tiny
	\begin{center}
		\caption{Physical and orbital characteristics of 
			exoplanets and their host stars}
		\label{T1}
		\begin{tabular}{@{}lccccccccc}
			\hline
			Name& M$_{p}$ & $\delta$ M$_{p}$ & $a$ & $\delta$ $a$ 
			& M$_{\star}$ & $\delta$ M$_{\star}$ & Spectral & T$_{\star}$ & $\delta$ T$_{\star}$  \\
			&(M$_{J}$) & (M$_{J}$) & (AU) & AU & (M$_{\odot}$) & (M$_{\odot}$) & Type & (Gyrs) & (Gyrs) \\ \hline
			
			24 Sex b & 1.990 & 0.380 & 1.333 & 0.009 & 1.540 & 0.080 & G5 & 2.70 & 0.40 \\
			24 Sex c & 0.860 & 0.220 & 2.080 & 0.020 & 1.540 & 0.080 & G5 & 2.70 & 0.40 \\
			47 Uma b & 2.530 & 0.060 & 2.100 & 0.020 & 1.030 & 0.050 & G0V & 7.40 & 1.90 \\
			47 Uma c & 0.540 & 0.073 & 3.600 & 0.100 & 1.030 & 0.050 & G0V & 7.40 & 1.90 \\
			47 Uma d & 1.640 & 0.480 & 11.600 & 2.900 & 1.030 & 0.050 & G0V & 7.40 & 1.90 \\
			61 Vir b & 0.016 & 0.001 & 0.050 & 0.000 & 0.950 & 0.030 & G5V & 8.96 & 3.00 \\
			61 Vir c & 0.057 & 0.003 & 0.217 & 0.000 & 0.950 & 0.030 & G5V & 8.96 & 3.00 \\
			61 Vir d & 0.072 & 0.008 & 0.476 & 0.001 & 0.950 & 0.030 & G5V & 8.96 & 3.00 \\
			CoRoT-12 b & 0.917 & 0.065 & 0.040 & 0.001 & 1.078 & 0.072 & G2V & 6.30 & 3.10 \\
			CoRoT-16 b & 0.535 & 0.085 & 0.061 & 0.001 & 1.098 & 0.078 & G5V & 6.73 & 2.80 \\
			CoRoT-17 b & 2.430 & 0.160 & 0.046 & 0.001 & 1.040 & 0.100 & G2V & 10.70 & 1.00 \\
			CoRoT-23 b & 2.800 & 0.250 & 0.047 & 0.004 & 1.140 & 0.080 & G0V & 7.20 & 1.00 \\
			CoRoT-25 b & 0.270 & 0.040 & 0.057 & 0.003 & 1.090 & 0.050 & G0V & 5.20 & 1.30 \\
			CoRoT-26 b & 0.520 & 0.050 & 0.052 & 0.001 & 1.090 & 0.060 & G8IV & 8.60 & 1.80 \\
			CoRoT-28 b & 0.484 & 0.087 & 0.059 & 0.003 & 1.010 & 0.140 & G8/9IV & 12.00 & 1.50 \\
			HAT-P-13 b & 0.850 & 0.038 & 0.042 & 0.001 & 1.220 & 0.100 & G4 & 5.00 & 0.80 \\
			HAT-P-15 b & 1.946 & 0.066 & 0.096 & 0.001 & 1.013 & 0.043 & G5 & 6.80 & 1.60 \\
			HAT-P-21 b & 4.063 & 0.161 & 0.049 & 0.001 & 0.947 & 0.042 & G3 & 10.20 & 2.50 \\ 
			HAT-P-22 b & 2.147 & 0.061 & 0.041 & 0.001 & 0.916 & 0.035 & G5 & 12.40 & 2.60 \\
			HAT-P-23 b & 2.090 & 0.110 & 0.023 & 0.000 & 1.130 & 0.050 & G5 & 4.00 & 1.00 \\
			HAT-P-28 b & 0.626 & 0.037 & 0.043 & 0.001 & 1.025 & 0.047 & G3 & 6.10 & 1.90 \\
			HAT-P-38 b & 0.267 & 0.020 & 0.052 & 0.001 & 0.886 & 0.044 & G & 10.10 & 4.80 \\
			HATS-11 b & 0.850 & 0.120 & 0.046 & 0.001 & 1.000 & 0.060 & G0 & 7.70 & 1.60 \\
			HATS-19 b & 0.427 & 0.071 & 0.058 & 0.001 & 1.303 & 0.830 & G0 & 3.94 & 0.50 \\
			HATS-25 b & 0.613 & 0.042 & 0.051 & 0.001 & 0.994 & 0.035 & G & 7.50 & 1.90 \\
			HATS-28 b & 0.672 & 0.087 & 0.041 & 0.001 & 0.929 & 0.036 & G & 6.20 & 2.80 \\
			HATS-29 b & 0.653 & 0.063 & 0.054 & 0.001 & 1.032 & 0.049 & G & 5.50 & 1.70 \\
			HATS-8 b & 0.138 & 0.019 & 0.046 & 0.001 & 1.056 & 0.037 & G2V & 5.10 & 1.70 \\
			HD 10180 c & 0.041 & - & 0.064 & 0.001 & 1.060 & 0.050 & G1V & 4.30 & 0.50 \\
			HD 10180 d & 0.036 & - & 0.128 & 0.002 & 1.060 & 0.050 & G1V & 4.30 & 0.50 \\ 
			HD 10180 e & 0.078 & - & 0.269 & 0.004 & 1.060 & 0.050 & G1V & 4.30 & 0.50 \\
			HD 10180 f & 0.075 & - & 0.492 & 0.007 & 1.060 & 0.050 & G1V & 4.30 & 0.50 \\
			HD 10180 g & 0.067 & - & 1.422 & 0.026 & 1.060 & 0.050 & G1V & 4.30 & 0.50 \\
			HD 10180 h & 0.202 & - & 3.400 & 0.110 & 1.060 & 0.050 & G1V & 4.30 & 0.50 \\
			HD 102365 b & 0.050 & 0.008 & 0.460 & 0.040 & 0.850 & - & G2V & 9.00 & 3.00 \\
			HD 104985 b & 8.300 & - & 0.950 & - & 1.600 & - & G9 III & 2.95 & 0.65 \\
			HD 106270 b & 11.000 & 0.800 & 4.300 & 0.400 & 1.320 & 0.092 & G5 & 4.30 & 0.60 \\
			HD 10697 b & 6.830 & 0.984 & 2.160 & 0.120 & 1.150 & 0.030 & G5 IV & 6.90 & 0.60 \\
			HD 109271 b & 0.054 & 0.004 & 0.079 & 0.001 & 1.047 & 0.024 & G5V & 7.30 & 1.20 \\
			HD 109271 c & 0.076 & 0.007 & 0.196 & 0.003 & 1.047 & 0.024 & G5V & 7.30 & 1.20 \\
			HD 109749 b & 0.280 & 0.016 & 0.063 & - & 1.200 & 0.100 & G3 IV & 10.30 & 2.90 \\
			HD 11506 b & 3.440 & 0.470 & 2.430 & 0.120 & 1.190 & 0.100 & G0V & 5.40 & 1.60 \\
			HD 11506 c & 0.820 & 0.500 & 0.639 & 0.017 & 1.190 & 0.100 & G0V & 5.40 & 1.60 \\
			HD 117207 b & 2.060 & - & 3.780 & - & 1.070 & - & G8VI/V & 6.68 & 2.20 \\
			HD 11755 b & 6.500 & 1.000 & 1.080 & 0.040 & 0.900 & 0.100 & G5 & 10.20 & 1.30 \\
			HD 12648 b & 2.900 & 0.400 & 0.540 & 0.020 & 1.200 & 0.100 & G5 & 4.50 & 1.00 \\
			HD 132406 b & 5.610 & - & 1.980 & - & 1.090 & 0.050 & G0V & 6.40 & 0.80 \\
			HD 134987 b & 1.590 & 0.020 & 0.810 & 0.020 & 1.070 & 0.080 & G5 V & 9.70 & 3.70 \\
			HD 134987 c & 0.820 & 0.030 & 5.800 & 0.500 & 1.070 & 0.080 & G5 V & 9.70 & 3.70 \\
			HD 136418 b & 2.000 & 0.100 & 1.320 & 0.030 & 1.330 & 0.090 & G5 & 4.00 & 1.00 \\
			HD 13931 b & 1.880 & 0.150 & 5.150 & 0.290 & 1.020 & 0.020 & G0 & 8.40 & 2.00 \\
			HD 14067 b & 7.800 & 0.700 & 3.400 & 0.100 & 2.400 & 0.200 & G9 III & 0.69 & 0.20 \\
			HD 149026 b & 0.357 & 0.011 & 0.042 & 0.000 & 1.300 & 0.100 & G0 IV & 2.00 & 0.80 \\
			HD 149143 b & 1.330 & - & 0.052 & - & 1.100 & 0.100 & G0 IV & 7.60 & 1.20 \\
			HD 154672 b & 5.020 & 0.170 & 0.600 & 0.170 & 1.060 & 0.090 & G3IV & 9.28 & 2.36 \\
			HD 16175 b & 4.770 & 0.370 & 2.148 & 0.076 & 1.350 & 0.090 & G0 & 5.30 & 1.00 \\
			HD 162004 b & 1.530 & 0.100 & 4.430 & 0.040 & 1.190 & 0.070 & G0V & 3.30 & 1.30 \\
			HD 16417 b & 0.069 & 0.007 & 0.140 & 0.010 & 1.180 & 0.040 & G1V & 4.30 & 0.80 \\
			HD 165155 b & 2.890 & 0.230 & 1.130 & 0.040 & 1.020 & 0.050 & G8V & 11.00 & 4.00 \\
			HD 168443 b & 7.659 & 0.097 & 0.293 & 0.001 & 0.995 & 0.019 & G5 & 9.80 & 1.00 \\
			HD 170469 b & 0.670 & - & 2.240 & - & 1.140 & 0.020 & G5IV & 6.70 & 1.10 \\
			HD 171028 b & 1.980 & - & 1.320 & - & 0.990 & 0.080 & G0 & 8.00 & 2.00 \\
			HD 17156 b & 3.195 & 0.033 & 0.162 & 0.002 & 1.275 & 0.018 & G0 & 3.38 & 0.47 \\
			HD 175607 b & 0.028 & 0.003 & - & - & 0.710 & 0.010 & G6V & 10.32 & 1.58 \\
			HD 187085 b & 0.750 & - & 2.050 & - & 1.220 & 0.100 & G0 V & 3.30 & 1.20 \\
			HD 18742 b & 2.700 & 0.300 & 1.920 & 0.050 & 1.600 & 0.110 & G9IV & 2.30 & 0.50 \\
			HD 188015 b & 1.260 & - & 1.190 & - & 1.090 & - & G5IV & 6.20 & 2.32 \\
			HD 20794 b & 0.008 & 0.001 & 0.120 & 0.002 & 0.850 & 0.040 & G8V & 14.00 & 5.00 \\
			HD 20794 c & 0.007 & 0.001 & 0.203 & 0.003 & 0.850 & 0.040 & G8V & 14.00 & 5.00 \\ 
			HD 20794 d & 0.015 & 0.002 & 0.349 & 0.006 & 0.850 & 0.040 & G8V & 14.00 & 5.00 \\
			HD 209458 b & 0.690 & 0.017 & 0.047 & 0.000 & 1.148 & 0.022 & G0V & 4.00 & 2.00 \\
			HD 212771 b & 2.300 & 0.400 & 1.220 & 0.300 & 1.150 & 0.080 & G8IV & 6.00 & 2.00 \\
			HD 219077 b & 10.390 & 0.090 & 6.220 & 0.090 & 1.050 & 0.020 & G8V & 8.90 & 0.30 \\
			HD 219828 b & 0.066 & 0.004 & 0.045 & - & 1.240 & - & G0IV & 5.80 & 1.20 \\	
				HD 221585 b & 1.610 & 0.140 & 2.306 & 0.081 & 1.190 & 0.120 & G8IV & 6.20 & 0.50 \\
				HD 222155 b & 1.900 & 0.530 & 5.100 & 0.700 & 1.130 & 0.110 & G2V & 8.20 & 0.70 \\
				HD 224693 b & 0.710 & - & 0.233 & - & 1.330 & 0.100 & G2IV & 2.00 & 0.50 \\
				HD 24040 b & 4.010 & 0.490 & 4.920 & 0.380 & 1.180 & - & G0 & 6.68 & 1.52 \\
				HD 30177 b & 8.070 & 0.120 & 3.580 & 0.010 & 1.050 & 0.080 & G8 V & 11.60 & 2.20 \\
				HD 30177 c & 3.000 & 0.300 & 6.990 & 0.420 & 1.050 & 0.080 & G8 V & 11.60 & 2.20 \\
				HD 34445 b & 0.790 & 0.070 & 2.070 & 0.020 & 1.070 & 0.020 & G0 & 8.50 & 2.00 \\
				HD 38529 b & 0.930 & 0.110 & 0.131 & 0.001 & 1.480 & 0.050 & G4 IV & 3.28 & 0.30 \\
				HD 4308 b & 0.040 & 0.005 & 0.118 & 0.009 & 0.850 & - & G5 V & 7.10 & 3.00 \\
				HD 4313 b & 2.300 & 0.200 & 1.190 & 0.030 & 1.720 & 0.120 & G5 D & 2.00 & 0.50 \\
				HD 43691 b & 2.490 & - & 0.240 & - & 1.380 & 0.050 & G0IV & 2.80& 0.80 \\
				HD 5319 b & 1.940 & - & 1.750 & - & 1.560 & 0.180 & G5IV & 2.40 & 0.68 \\
				HD 5319 c & 1.150 & 0.080 & 2.071 & 0.013 & 1.560 & 0.180 & G5IV & 2.40 & 0.68 \\
				\hline
		\end{tabular}
		\end {center}
	\end{table*}
	
	\newpage
	
	\begin{table*}
		\tiny
		\begin{center}
			\centerline{Table 1 continued}
			\begin{tabular}{@{}lccccccccc}
				\hline
				Name& M$_{p}$ & $\delta$ M$_{p}$ & $a$ & $\delta$ $a$ 
				& M$_{\star}$ & $\delta$ M$_{\star}$ & Spectral & T$_{\star}$ & $\delta$ T$_{\star}$  \\
				&(M$_{J}$) & (M$_{J}$) & (AU) & AU & (M$_{\odot}$) & (M$_{\odot}$) & Type & (Gyrs) & (Gyrs) \\ \hline
				
				HD 72659 b & 3.150 & 0.140 & 4.740 & 0.080 & 0.950 & 2.000 & G2 V & 6.50 & 1.50 \\
				HD 72892 b & 5.450 & 0.370 & 0.228 & 0.008 & 1.020 & 0.050 & G5V & 8.00 & 3.00 \\
				HD 74156 b & 1.778 & 0.020 & 0.291 & 0.003 & 1.240 & 0.040 & G1V & 3.70 & 0.40 \\
				HD 74156 c & 7.997 & 0.095 & 3.820 & 0.044 & 1.240 & 0.040 & G1V & 3.70 & 0.40 \\
				HD 75898 b & 2.510 & - & 1.190 & - & 1.280 & 0.130 & G0 & 3.80 & 0.80 \\
				HD 82886 b & 1.300 & 0.100 & 1.650 & 0.060 & 1.060 & 0.074 & G0 & 7.00 & 2.00 \\
				HD 9174 b & 1.110 & 0.140 & 2.200 & 0.090 & 1.030 & 0.050 & G8IV & 9.00 & 3.00 \\
				HD 96167 b & 0.680 & 0.180 & 1.300 & 0.070 & 1.310 & 0.090 & G5D & 3.80 & 1.00 \\
				HIP 11915 & 0.990 & 0.060 & 4.800 & 0.100 & 1.000 & - & G5V & 4.00 & 0.60 \\
				HIP 68468 b & 0.011 & - & 0.029 & - & 1.050 & 0.010 & G3V & 5.90 & 0.40 \\
				HIP 68468 c & 0.094 & - & 0.665 & - & 1.050 & 0.010 & G3V & 5.90 & 0.40 \\
				K2-24 b & 0.066 & 0.017 & 0.154 & 0.002 & 1.120 & 0.050 & G9V & 5.00 & 1.80 \\
				K2-24 c & 0.085 & 0.022 & 0.247 & 0.004 & 1.120 & 0.050 & G9V & 5.00 & 1.80 \\
				K2-60 b & 0.426 & 0.037 & 0.045 & 0.003 & 0.970 & 0.070 & G4V & 10.00 & 3.00 \\
				K2-99 b & 0.970 & 0.090 & 0.159 & 0.006 & 1.600 & 0.100 & G0 IV & 2.40 & 0.60 \\
				KELT-8 b & 0.874 & 0.066 & 0.045 & 0.009 & 1.211 & 0.066 & G2V & 5.40 & 0.50 \\
				Kepler-10 b & 0.010 & 0.001 & 0.016 & 0.000 & 0.910 & 0.021 & G & 10.60 & 1.50 \\
				Kepler-10 c & 0.054 & 0.005 & 0.241 & 0.001 & 0.910 & 0.021 & G & 10.60 & 1.50 \\
				Kepler-11 b & 0.005 & 0.003 & 0.091 & 0.003 & 0.950 & 0.100 & G & 8.00 & 2.00 \\
				Kepler-11 c & 0.009 & 0.005 & 0.106 & 0.004 & 0.950 & 0.100 & G & 8.00 & 2.00 \\
				Kepler-11 d & 0.022 & 0.004 & 0.159 & 0.005 & 0.950 & 0.100 & G & 8.00 & 2.00 \\
				Kepler-11 e & 0.030 & 0.006 & 0.194 & 0.007 & 0.950 & 0.100 & G & 8.00 & 2.00 \\
				Kepler-11 f & 0.006 & 0.002 & 0.250 & 0.009 & 0.950 & 0.100 & G & 8.00 & 2.00 \\
				Kepler-11 g & 0.950 & 0.950 & 0.462 & 0.016 & 0.950 & 0.100 & G & 8.00 & 2.00 \\
				Kepler-12 b & 0.431 & 0.041 & 0.055 & 0.001 & 1.166 & 0.054 & G0 & 4.00 & 0.40 \\
				Kepler-20 b & 0.030 & 0.004 & 0.046 & 0.001 & 0.912 & 0.035 & G8 & 8.80 & 2.70 \\
				Kepler-20 c & 0.040 & 0.007 & 0.094 & 0.002 & 0.912 & 0.035 & G8 & 8.80 & 2.70 \\
				Kepler-20 d & 0.031 & 0.011 & 0.350 & 0.010 & 0.912 & 0.035 & G8 & 8.80 & 2.70 \\
				Kepler-20 e & 0.009 & - & 0.063 & 0.001 & 0.912 & 0.035 & G8 & 8.80 & 2.70 \\
				Kepler-20 f & 0.045 & - & 0.139 & 0.003 & 0.912 & 0.035 & G8 & 8.80 & 2.70 \\
				Kepler-20 g & 0.062 & 0.011 & 0.205 & 0.002 & 0.912 & 0.035 & G8 & 8.80 & 2.70 \\
				Kepler-4 b & 0.082 & 0.012 & 0.045 & 0.001 & 1.223 & 0.091 & G0 & 4.50 & 1.50 \\
				Kepler-41  b & 0.490 & 0.090 & 0.029 & 0.001 & 0.940 & 0.090 & G2V & 7.40 & 3.40 \\
				Kepler-412 b & 0.939 & 0.085 & 0.029 & 0.001 & 1.167 & 0.091 & G3V & 5.10 & 1.70 \\
				Kepler-43 b & 3.230 & 0.190 & 0.044 & 0.001 & 1.320 & 0.090 & G0V/G0IV & 2.80 & 0.80 \\
				Kepler-44 b & 1.020 & 0.070 & 0.045 & 0.001 & 1.190 & 0.100 & G2IV & 6.95 & 1.70 \\
				Kepler-66 b & 0.310 & 0.070 & 0.135 & 0.001 & 1.038 & 0.044 & GOV & 1.00 & 0.17 \\
				Kepler-67 b & 0.310 & 0.060 & 0.117 & 0.001 & 0.865 & 0.034 & G9V & 1.00 & 0.17 \\
				Kepler-75 b & 9.900 & 0.500 & 0.080 & 0.005 & 0.880 & 0.060 & G8V & 6.00 & 3.00 \\
				Kepler-77 b & 0.430 & 0.032 & 0.045 & 0.001 & 0.950 & 0.040 & G5V & 7.50 & 2.00 \\
				Pr 0211 b & 1.880 & 0.020 & 0.031 & 0.001 & 0.935 & 0.013 & G9 & 0.79 & 0.03 \\
				Pr 0211 c & 7.950 & 0.250 & 5.800 & 1.400 & 0.935 & 0.013 & G9 & 0.79 & 0.03 \\
				WASP-102 b & 0.624 & 0.045 & 0.040 & 0.000 & 1.167 & 0.035 & G0 & 0.60 & 0.30 \\
				WASP-110 b & 0.515 & 0.064 & 0.045 & 0.001 & 0.892 & 0.072 & G9 & 8.60 & 3.50 \\
				WASP-112 b & 0.880 & 0.120 & 0.038 & 0.001 & 0.807 & 0.073 & G6 & 10.60 & 3.00 \\
				WASP-119 b & 1.230 & 0.083 & 0.036 & 0.001 & 1.020 & 0.060 & G5 & 8.00 & 2.50 \\
				WASP-12 b & 1.404 & 0.099 & 0.022 & 0.001 & 1.350 & 0.140 & G0 & 1.70 & 0.80 \\
				WASP-126 b & 0.280 & 0.040 & 0.044 & 0.001 & 1.120 & 0.060 & G2 & 6.40 & 1.60 \\
				WASP-127 b & 0.180 & 0.020 & 0.052 & 0.001 & 1.080 & 0.030 & G5 & 11.41 & 1.80 \\ 
				WASP-133 b & 1.160 & 0.090 & 0.034 & 0.001 & 1.160 & 0.080 & G4 & 6.80 & 1.80 \\
				WASP-157 b & 0.576 & 0.093 & 0.052 & 0.001 & 1.260 & 0.120 & G2V & 1.00 & 0.30 \\
				WASP-19 b & 1.114 & 0.040 & 0.016 & 0.000 & 0.904 & 0.045 & G8V & 11.50 & 2.70 \\
				WASP-26 b & 1.028 & 0.021 & 0.039 & 0.000 & 1.120 & 0.030 & G0 & 6.00 & 2.00 \\
				WASP-37 b & 1.800 & 0.170 & 0.043 & 0.001 & 0.849 & 0.040 & G2 & 11.00 & 4.00 \\
				WASP-46 b & 2.101 & 0.073 & 0.024 & 0.000 & 0.956 & 0.034 & G6V & 1.40 & 0.60 \\
				WASP-5 b & 1.637 & 0.082 & 0.027 & 0.000 & 1.000 & 0.060 & G5 & 3.00 & 1.40 \\ 
				WASP-8 b & 2.244 & 0.093 & 0.080 & 0.001 & 1.033 & 0.050 & G6 & 4.00 & 1.00 \\
				WASP-95 b & 1.130 & 0.040 & 0.034 & 0.001 & 1.110 & 0.090 & G2 & 2.40 & 1.00 \\
				XO-1 b & 0.900 & 0.070 & 0.048 & 0.000 & 1.000 & 0.030 & G1V & 4.50 & 2.00 \\
				XO-5 b & 1.077 & 0.037 & 0.048 & 0.000 & 0.880 & 0.030 & G8V & 8.50 & 0.80 \\		
				\hline
			\end{tabular}
			\end {center}
		\end{table*}
				
		\begin{table*}
					\tiny
					\caption{Rate of Mass loss of stars with planets}
					\fontsize{6pt}{6pt}
					\label{T2}
					\begin{tabular}{@{}lcccc}
						\hline
						& Mass loss obtained & 
						Mass loss obtained & 
						Mass loss obtained \\
						& from host star data & from observed data & from empirical relation \\ \hline
						Rate of mass loss $\frac{dM}{dt}$ & 
						$c_1(\frac{M_{\star}}{M_{\odot}})^{3.788\pm0.190}$ &
						$c_2(\frac{M_{\star}}{M_{\odot}})^{3.974\pm1.394}$ & 
						$10^{-8.158}[\frac{(\frac{L_{\star}}{L_{\odot}})^{1.768}}{({T_{eff}})^{1.676}}]$ \\ \\
						
						Sun$^{\prime}$s rate of mass loss & 1.361 x $10^{-11} M_{\odot}$/yr & 8.090 x $10^{-13}M_{\odot}$/yr & 3.4435 x $10^{-15}M_{\odot}$/yr \\ \\
						
						Initial $M_{\odot}$ & $1.061\pm0.006$ $M_{\odot}$ & $1.003\pm0.006$ $M_{\odot}$
						& $1.000015$ $M_{\odot}$ \\				
						\hline 
					\end{tabular}
					
					where $c_1$=(0.062$\pm$0.003), $c_2$=(0.003$\pm$0.002), $M_{\star}$ - mass of the star, $\tau_{\odot}$ - age of Sun. \\
					Note: (i) The results from second column are obtained from the exoplanetary dataset as explained in section \ref{3.1},
					results from third column are obtained from observed mass loss data from the \cite{Cranmer} as explained in section \ref{3.2},
					and the results from the last column are obtained by using empirical formula from \cite{Jager}.
		\end{table*}
		
		\begin{table*}
					\tiny
					\begin{center}
						\caption{Relationship between host stars initial mass and planetary mass}
						\label{T3}
						\begin{tabular}{@{}lccc}
							\hline
							Corrections applied & $(\frac{M_{p}}{M_{J}})$ versus initial $(\frac{M_{\star}}{M_{\odot}})_{ini}$ relationships
							& Missing planetary mass \\ & & in the vicinity of Sun \\ \hline
							Host star data correction & $log \left(\frac{M_{p}}{M_{J}}\right)=(-0.191\pm0.045)+(2.366\pm0.220)log\left(\frac{M_{\star}}{M_{\odot}}\right)_{ini}$ & 
							0.738$\pm$0.172 $M_{J}$ \\
							Observed mass loss correction & $log\left(\frac{M_{p}}{M_{J}}\right)=(-0.116\pm0.041)+(2.560\pm0.223)log\left(\frac{M_{\star}}{M_{\odot}}\right)_{ini}$ &
							0.887$\pm$0.191 $M_{J}$ \\
							Empirical mass loss correction & $log\left(\frac{M_{p}}{M_{J}}\right)=(-0.102\pm0.041)+(2.522\pm0.233)log\left(\frac{M_{\star}}{M_{\odot}}\right)_{ini}$
							& 0.914$\pm$0.198 $M_{J}$ \\
							\\ \hline
							Average initial planetary mass & 0.846$\pm$0.187 $M_{J}$ & \\ \hline
						\end{tabular} 
					\end{center}
		\end{table*}

\newpage
\bibliography{ref.bib}

\end{document}